\documentclass[12pt]{article}
\usepackage{bbm,latexsym}
\usepackage{epic}
\newcommand{\be}{\begin{equation}}
\newcommand{\ee}{\end{equation}}
\newcommand{\ba}{\begin{eqnarray}}
\newcommand{\ea}{\end{eqnarray}}
\newcommand{\no}{\nonumber\\}
\newcommand{\kk}[2]{\kappa^{(#1#2)}}
\newcommand{\lesssim}{\:\mbox{\raisebox{-3pt}{$\stackrel%
{\displaystyle <}{\sim}$}}\:}
\newcommand{\gtrsim}{\:\mbox{\raisebox{-3pt}{$\stackrel%
{\displaystyle >}{\sim}$}}\:}
\def\e{\epsilon}
\begin{document}
\title{\normalsize \hfill UWThPh-2004-23 \\[1cm] \LARGE
Renormalization of the neutrino mass operators \\
in the multi-Higgs-doublet Standard Model}
\author{Walter Grimus\thanks{E-mail: walter.grimus@univie.ac.at} \\
\setcounter{footnote}{6}
\small Institut f\"ur Theoretische Physik, Universit\"at Wien \\
\small Boltzmanngasse 5, A--1090 Wien, Austria \\*[3.6mm]
Lu\'\i s Lavoura\thanks{E-mail: balio@cfif.ist.utl.pt} \\
\small Universidade T\'ecnica de Lisboa
and Centro de F\'\i sica Te\'orica de Part\'\i culas \\
\small Instituto Superior T\'ecnico,
P--1049-001 Lisboa, Portugal \\*[4.6mm] }

\date{20 September 2004}

\maketitle

\begin{abstract}
We derive the renormalization group equations (RGE)
for the flavour coupling matrices
of the effective dimension-five operators
which yield Majorana neutrino masses
in the multi-Higgs-doublet Standard Model;
in particular,
we consider the case where two different scalar doublets
occur in those operators.
We also write down the RGE for the scalar-potential quartic couplings
and for the Yukawa couplings of that model,
in the absence of quarks.
As an application of the RGE,
we consider two models which,
based on a $\mu$--$\tau$ interchange symmetry,
predict maximal atmospheric neutrino mixing,
together with $U_{e3} = 0$,
at the seesaw scale.
We estimate the change of those predictions
due to the evolution of the coupling matrices
of the effective mass operators
from the seesaw scale down to the electroweak scale.
We derive an upper bound on that change,
thereby finding that the radiative corrections
to those predictions are in general negligible. 
\end{abstract}

\newpage


\section{Introduction}

The Standard Model (SM) in the strict sense,
i.e.\ without right-handed neutrino singlets,
forbids neutrino masses.
However,
it was noticed a long time ago \cite{weinberg} that,
if one allows for lepton-number nonconservation,
then one can construct,
with the SM multiplets,
operators of dimension higher than four
which give Majorana masses to the neutrinos.
The lowest-dimensional such operators
have dimension five and contain
two left-handed lepton doublets and two Higgs doublets;
those operators can be thought of
as arising from the seesaw mechanism \cite{seesaw}
after one has integrated out the right-handed neutrino
singlets.\footnote{The effect of the dimension-six operators
which also arise from the seesaw mechanism
has been studied in~\cite{gavela}.}
Under the assumption that the SM is valid
up to the seesaw scale $m_R$,
the renormalization group evolution
of the dimension-five neutrino mass operators
from $m_R$ down to the electroweak scale,
as represented for instance by the $Z^0$ mass $m_Z$,
can be determined;
the evolution equations have been computed
in the SM and in its minimal supersymmetric
extension~\cite{chankowski,babu,drees1,drees2}
(for a review see~\cite{pokorski}).
This is an important issue
in view of testing mechanisms and symmetries
for explaining the neutrino masses and the lepton mixing angles,
since such mechanisms and symmetries
are usually operative or imposed at the seesaw scale,
while the measurements are effected at the electroweak scale.
(For the experimental and theoretical status
of neutrino masses and lepton mixing see,
for instance,
\cite{maltoni} and~\cite{altarelli},
respectively.)

In this paper we extend the existing results
for the SM renormalization group equations (RGE)
to the case of an arbitrary number of Higgs doublets. 
In particular,
we focus on dimension-five neutrino mass operators
which contain two \emph{different} Higgs doublets;
indeed,
to our knowledge,
that case has not yet been treated in the literature.
The reason to consider the multi-Higgs-doublet SM is that,
within that framework,
several models have been produced in recent years which predict,
for instance,
lepton mixing angles $\theta_{13} = 0$
and $\theta_{23} = \pi / 4$~\cite{GLZ2,GLD4},
or $\theta_{13} = 0$ alone~\cite{low,softD4},
or $\theta_{23} = \pi / 4$
and $\delta = \pi / 2$~\cite{CP}.\footnote{The predictions
$\theta_{23} = \pi / 4$ and $\delta = \pi / 2$ have
first been obtained in a supersymmetric extension of the SM~\cite{ma}.}
(See~\cite{maltoni,altarelli},
for instance,
for the definition of the lepton mixing angles.)
Those predictions usually hold at the seesaw scale and,
in order to compare them with experiment,
one needs to know the corresponding corrections
at the electroweak scale. 

In Sect.~\ref{RGEs} we display the Lagrangian
of the multi-Higgs-doublet SM,
without quarks but with dimension-five neutrino mass operators,
and present the RGE for the couplings of that Lagrangian.
In Sect.~\ref{models} we discuss the specific RGE
for the models,
referred to above,
which predict $\theta_{13} = 0$ and $\theta_{23} = \pi / 4$
at the seesaw scale.
In Sect.~\ref{sectionpredictions} we show explicitly
how those sessaw-scale predictions arise,
and how they may be changed by the renormalization group evolution.
In Sect.~\ref{sectionupper} we derive an upper bound
on the effect of that evolution.
A short summary of our main results
is provided in Sect.~\ref{summary}. 
An appendix contains some details
of the calculation of the beta functions
for the neutrino mass operators.


\section{General case} \label{RGEs}

\subsection{The model}

We consider the SM with $n_H$ Higgs doublets $\phi_i$
($i = 1, 2, \ldots, n_H$)
with weak hypercharge $1/2$.
The $SU(2)$ gauge coupling constant is denoted $g$
while the $U(1)$ gauge coupling constant
(with the above normalization for the weak hypercharge)
is denoted $g^\prime$.
The scalar potential $V$ has the form 
\be
V = \mbox{quadratic terms}
+ \sum_{i,j,k,l = 1}^{n_H} \lambda_{ijkl}
\left( \phi_i^\dagger \phi_j \right)
\left( \phi_k^\dagger \phi_l \right),
\label{pot}
\ee
where the dimensionless couplings $\lambda_{ijkl}$ satisfy
\be
\lambda_{ijkl} = \lambda_{klij} = \lambda_{jilk}^\ast.
\label{ident}
\ee
The lepton Yukawa Lagrangian $\mathcal{L}_{\mathrm{Y} \ell}$ is
\be
\mathcal{L}_{\mathrm{Y} \ell} =
- \sum_{i=1}^{n_H} \left(
\bar \ell_\mathrm{R} \phi_i^\dagger Y_i D_\mathrm{L}
+ \bar D_\mathrm{L} Y_i^\dagger \phi_i \ell_\mathrm{R} \right),
\label{ly}
\ee
where $D_\mathrm{L}$ denotes the left-handed lepton doublets
and $\ell_\mathrm{R}$ the right-handed charged-lepton singlets.
We have defined the dimensionless flavour coupling matrices $Y_i$
in the same way as~\cite{drees1,drees2}.
Note that in this paper we do not use the summation convention.

The effective dimension-five neutrino mass operators
are defined as
\be
\mathcal{O}_{ij} = \sum_{\alpha, \beta = e, \mu, \tau}\
\! \sum_{a, b, c, d = 1}^2
\left( D_{\mathrm{L} \alpha a}^T \kk ij_{\alpha \beta}
C^{-1} D_{\mathrm{L} \beta c} \right)
\left( \varepsilon^{ab} \phi_{ib} \right)
\left( \varepsilon^{cd} \phi_{jd} \right),
\label{Op}
\ee
where,
contrary to what we had done in~(\ref{pot}) and~(\ref{ly}),
we have made explicit both the flavour and gauge-$SU(2)$ indices,
and the summations thereover.
In~(\ref{Op}),
$C$ is the Dirac--Pauli charge-conjugation matrix;
$\alpha$ and $\beta$ are flavour indices;
$a$, $b$, $c$, and $d$ are $SU(2)$ indices;
and $\varepsilon$ is the antisymmetric $2 \times 2$ matrix,
with $\epsilon^{12} = 1$.
The flavour coupling matrices $\kk ij$ in~(\ref{Op})
have dimension $-1$ and satisfy
\be
\kk ij_{\alpha \beta} = \kk ji_{\beta \alpha},
\ \mbox{i.e.} \
\kk ij = {\kk ji}^T.
\label{transp}
\ee
%


\subsection{The RGE}

The RGE are first-order differential equations
which give the evolution of the couplings of a model
relative to $t = \ln{\mu}$,
where $\mu$ is the mass parameter
used in the regularization of ultraviolet-divergent integrals;
the basic equation is
\be
\int \! \frac{\mathrm{d}^4 k}{\left( 2 \pi \right)^4}\, 
\frac{1}{\left( k^2 - m^2 \right)
\left[ \left( k+p \right)^2 - {m^\prime}^2 \right]}
= \frac{i t}{8 \pi^2} + \cdots,
\ee
where $p$ is a typical momentum and $m$,
$m^\prime$ are typical masses
appearing in the loop integral;
the dots represent 
either $\mu$-independent terms
or terms which disappear
in the limit that regularizes the integral.
We have computed,
at the one-loop level,
the RGE for the model outlined in the previous subsection.
For the RGE of the coupling matrices of the effective mass operators 
we have found
\ba
16 \pi^2\, \frac{\mathrm{d} \kappa^{(ij)}}{\mathrm{d}t} &=&
- 3 g^2 \kappa^{(ij)}
+ 4 \sum_{k, l = 1}^{n_H} \lambda_{kilj} \kappa^{(kl)}
+ \sum_{k=1}^{n_H} \left[ T_{ki} \kk kj
+ T_{kj} \kk ik \right] + \kk ij P + P^T \kk ij
\no & &
+ 2\, \sum_{k=1}^{n_H} \left\{
\kk kj Y_i^\dagger Y_k
- \left[ \kk ik + \kk ki \right] Y_j^\dagger Y_k
\right. \no & & \left.
+ Y_k^T Y_j^\ast \kk ik  
- Y_k^T Y_i^\ast \left[ \kk kj + \kk jk \right]
\right\},
\label{RGEkappa} 
\ea
where
\ba
T_{ij} &:=& \mbox{tr} \left( Y_i Y_j^\dagger \right),
\label{Tdefinition}
\\
P &:=& \frac{1}{2}\, \sum_{k=1}^{n_H} Y_k^\dagger Y_k.
\label{Pdefinition}
\ea
The third line of~(\ref{RGEkappa}) is obtained from the second line
through the interchange $i \leftrightarrow j$
together with transposition,
in agreement with~(\ref{transp}).
Our result~(\ref{RGEkappa}) coincides,
when $i=j$,
with the result given in~\cite{drees1,drees2};
it generalizes that result for the case $i \neq j$.
Note that,
for the sake of simplicity,
in the present paper we dismiss quarks;
in general,
one would have to add to~(\ref{Tdefinition})
analogous trace terms featuring the Yukawa-coupling matrices
of the Higgs doublets $\phi_i$ and $\phi_j$ to the up and down quarks,
multiplied by a colour factor 3---see,
for instance,
\cite{babu}.

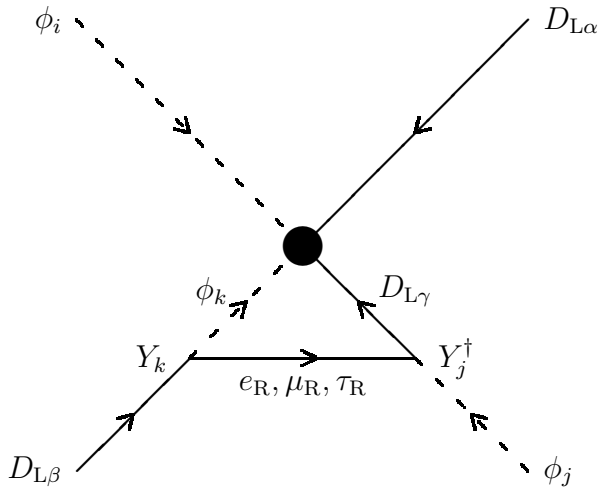
\begin{figure}[t]
\setlength{\unitlength}{1cm}
\begin{center}
\begin{picture}(8,8)
\put(4,4){\circle*{5}}
\thicklines
\dashline{0.15}(4,4)(1,7)
\dashline{0.15}(5.5,2.5)(7,1)
\dashline{0.15}(4,4)(2.5,2.5)
\drawline(4,4)(5.5,2.5)
\drawline(4,4)(7,7)
\drawline(2.5,2.5)(1,1)
\drawline(2.5,2.5)(5.5,2.5)
\drawline(2.3,5.55)(2.5,5.5) \drawline(2.45,5.7)(2.5,5.5)
\drawline(6.45,1.7)(6.25,1.75) \drawline(6.3,1.55)(6.25,1.75)
\drawline(3.05,3.2)(3.25,3.25) \drawline(3.2,3.05)(3.25,3.25)
\drawline(4.8,3.05)(4.75,3.25) \drawline(4.95,3.2)(4.75,3.25)
\drawline(5.55,5.7)(5.5,5.5) \drawline(5.7,5.55)(5.5,5.5)
\drawline(1.55,1.7)(1.75,1.75) \drawline(1.7,1.55)(1.75,1.75)
\drawline(4,2.6)(4.2,2.5) \drawline(4,2.4)(4.2,2.5)
\put(0.8,7){\makebox(0,0)[r]{$\phi_i$}}
\put(7.2,7){\makebox(0,0)[l]{$D_{\mathrm{L} \alpha}$}}
\put(7.2,1){\makebox(0,0)[l]{$\phi_j$}}
\put(0.8,1){\makebox(0,0)[r]{$D_{\mathrm{L} \beta}$}}
\put(4.0,2){\makebox(0,0)[b]{$e_\mathrm{R},
\mu_\mathrm{R}, \tau_\mathrm{R}$}}
\put(5.0,3.2){\makebox(0,0)[lb]{$D_{\mathrm{L} \gamma}$}}
\put(3.0,3.2){\makebox(0,0)[rb]{$\phi_k$}}
\put(2.2,2.5){\makebox(0,0)[r]{$Y_k$}}
\put(5.8,2.5){\makebox(0,0)[l]{$Y_j^\dagger$}}
\end{picture}
\end{center}
\caption{A typical vertex correction
in the renormalization of the operator $\mathcal{O}_{ij}$.
The relevant Yukawa-coupling matrices are indicated.
\label{fig}}
\end{figure}

The terms in the second and third lines of~(\ref{RGEkappa})
arise from diagrams like the one in figure~\ref{fig}.
We dwell on the explicit derivation of those terms in the appendix.

In order to solve the RGE for the effective neutrino mass operators
one also needs the RGE for the other couplings
occurring in~(\ref{RGEkappa}).
The general RGE for an arbitrary renormalizable gauge field theory
have been derived in~\cite{li,machacek} at the one- and two-loop levels,
respectively.
It is convenient to have the results of~\cite{li}
specialized to the case of the multi-Higgs-doublet SM.
We have found that
\ba
16 \pi^2\, \frac{\mathrm{d} \lambda_{ijkl}}{\mathrm{d}t} &=&
4 \sum_{m,n = 1}^{n_H} \left(
2\, \lambda_{ijmn} \lambda_{nmkl} + \lambda_{ijmn} \lambda_{kmnl}
+ \lambda_{imnj} \lambda_{mnkl}
\right. \no & & \left.
+ \lambda_{imkn} \lambda_{mjnl}
+ \lambda_{mjkn} \lambda_{imnl}
\right)
- \left( 9 g^2 + 3 {g^\prime}^2 \right) \lambda_{ijkl}
\no & &
+ \frac{9 g^4 + 3 {g^\prime}^4}{8}\, \delta_{ij} \delta_{kl}
+ \frac{3 g^2 {g^\prime}^2}{4}
\left( 2 \delta_{il} \delta_{kj} - \delta_{ij} \delta_{kl} \right)
\no & &
+ \sum_{m=1}^{n_H} \left( T_{mj} \lambda_{imkl} + 
T_{ml} \lambda_{ijkm} + T_{im} \lambda_{mjkl} + 
T_{km} \lambda_{ijml} \right) 
\no & &
- 2\, \mbox{tr} \left( Y_i Y_j^\dagger Y_k Y_l^\dagger \right),
\\ \label{yukawa}
16 \pi^2\, \frac{\mathrm{d} Y_i}{\mathrm{d}t} &=&
\sum_{k=1}^{n_H}
\left( T_{ik} Y_k + Y_k Y_k^\dagger Y_i
+ \frac{1}{2}\, Y_i Y_k^\dagger Y_k \right)
- \frac{9 g^2 + 15  {g^\prime}^2}{4}\,Y_i.
\ea
It is well known that the RGE for $g$ and $g^\prime$ are
\ba
16 \pi^2\, \frac{\mathrm{d} g}{\mathrm{d}t} &=&
\left( - \frac{22}{3} + \frac{4 N}{3} + \frac{n_H}{6} \right) g^3,
\\ 
16 \pi^2\, \frac{\mathrm{d} g^\prime}{\mathrm{d}t} &=&
\left( \frac{20 N}{9} + \frac{n_H}{6} \right) {g^\prime}^3,
\ea
where $N=3$ is the number of fermion families.


\section{Application of the RGE to two particular models}
\label{models}

\subsection{The $\mathbbm{Z}_2$ and $D_4$ models}

We now apply the general RGE derived in Sect.~\ref{RGEs}
to the so-called $\mathbbm{Z}_2$~\cite{GLZ2}
and $D_4$~\cite{GLD4} models---for
a review see~\cite{ustron03}.
Those models predict
\be
\begin{array}{rcl}
\theta_{13} &=& 0, \\
\theta_{23} &=& \pi / 4
\end{array}
\label{angles}
\ee
at the seesaw scale and are,
in what regards the practical application of the RGE,
identical.
They both have three Higgs doublets $\phi_1$,
$\phi_2$,
and $\phi_3$.
Below the seesaw scale the structure of both models
is dictated by the symmetries
\ba
\mathbbm{Z}_2^{\mathrm{(aux)}}: & &
e_\mathrm{R} \to - e_\mathrm{R},\
\phi_1 \to - \phi_1,
\label{Z2aux} \\
\mathbbm{Z}_2^{\mathrm{(tr)}}: & &
D_{\mathrm{L} \mu} \leftrightarrow D_{\mathrm{L} \tau},\
\mu_\mathrm{R} \leftrightarrow \tau_\mathrm{R},\
\phi_3 \to - \phi_3.
\label{Z2tr}
\ea
These two symmetries hold
in between the seesaw (high) scale $m_R$ 
and the electroweak (low) scale $m_Z$.
Indeed,
they are broken only spontaneously,
by the vacuum expectation values (VEVs) of $\phi_1^0$ and $\phi_3^0$,
respectively,
at the low scale.
Because of the symmetries in~(\ref{Z2aux}) and~(\ref{Z2tr}),
the Higgs potential is given by
\ba
V &=& \mbox{quadratic\ terms} + 
\sum_{i = 1}^3 \lambda_i \left( \phi_i^\dagger \phi_i \right)^2
\no & &
+ \lambda_4 \left( \phi_1^\dagger \phi_1 \right)
\left( \phi_2^\dagger \phi_2 \right)
+ \lambda_5 \left( \phi_1^\dagger \phi_1 \right)
\left( \phi_3^\dagger \phi_3 \right)
+ \lambda_6 \left( \phi_2^\dagger \phi_2 \right)
\left( \phi_3^\dagger \phi_3 \right)
\no & &
+ \lambda_7 \left( \phi_1^\dagger \phi_2 \right)
\left( \phi_2^\dagger \phi_1 \right)
+ \lambda_8 \left( \phi_1^\dagger \phi_3 \right)
\left( \phi_3^\dagger \phi_1 \right)
+ \lambda_9 \left( \phi_2^\dagger \phi_3 \right)
\left( \phi_3^\dagger \phi_2 \right)
\no & &
+ \left[ \lambda_{10} \left( \phi_1^\dagger \phi_2 \right)^2
+ \lambda_{11} \left( \phi_1^\dagger \phi_3 \right)^2
+ \lambda_{12} \left( \phi_2^\dagger \phi_3 \right)^2
+ \mbox{H.c.} \right],
\label{V}
\ea
where $\lambda_{10}$,
$\lambda_{11}$,
and $\lambda_{12}$ are the only non-real quartic couplings.
Comparing~(\ref{V}) with~(\ref{pot}) and~(\ref{ident}),
we arrive at the identifications 
$\lambda_{iiii} = \lambda_i$ (for $i= 1,2,3$), 
$\lambda_{1122} = \lambda_{2211} = \lambda_4/2$, 
$\lambda_{1221} = \lambda_{2112} = \lambda_7/2$, 
$\lambda_{1212} = \lambda_{2121}^* = \lambda_{10}$,
and so on.

The $\mathbbm{Z}_2$ and $D_4$ models have three other symmetries,
the family-lepton-number symmetries $L_\alpha$,
which are broken at the seesaw scale---softly in the $\mathbbm{Z}_2$ model,
spontaneously in the $D_4$ model.
Because of the symmetries in~(\ref{Z2aux}) and~(\ref{Z2tr}),
and also because of the symmetries $L_\alpha$---which remain valid
for the quartic couplings of the light fields
below the seesaw scale---the lepton Yukawa Lagrangian is
\be
\mathcal{L}_{\mathrm{Y} \ell} =
- y_3 \bar D_{\mathrm{L} e} e_\mathrm{R} \phi_1
- y_4 \left( \bar D_{\mathrm{L} \mu} \mu_\mathrm{R}
+ \bar D_{\mathrm{L} \tau} \tau_\mathrm{R} \right) \phi_2
- y_5 \left( \bar D_{\mathrm{L} \mu} \mu_\mathrm{R}
- \bar D_{\mathrm{L} \tau} \tau_\mathrm{R} \right) \phi_3
+ \mbox{H.c.}
\label{lyl}
\ee
(The coupling constants $y_{1,2}$
occur in the Yukawa interactions
of the right-handed neutrino singlets~\cite{ustron03}
and are thus of no concern here.)
Comparing~(\ref{lyl}) with~(\ref{ly}),
we see that the Yukawa-coupling matrices are
\ba
Y_1 &=& \mathrm{diag} \left( y_3^\ast,\ 0,\ 0 \right), \no
Y_2 &=& \mathrm{diag} \left( 0,\ y_4^\ast,\ y_4^\ast \right),
\label{ouryukawas} \\
Y_3 &=& \mathrm{diag} \left( 0,\ y_5^\ast,\ - y_5^\ast \right). \nonumber
\ea
Hence,
from~(\ref{Tdefinition}),
\ba
T_{11} &=& \left| y_3 \right|^2, \\
T_{22} &=& 2 \left| y_4 \right|^2, \\
T_{33} &=& 2 \left| y_5 \right|^2,
\ea
and the $T_{ij}$ with $i \neq j$ vanish,
a fact which simplifies considerably the RGE in this particular case.

As emphasized before,
the symmetries~(\ref{Z2aux}) and~(\ref{Z2tr})
are broken only at the electroweak scale.
The validity of the symmetry $\mathbbm{Z}_2^{\mathrm{(aux)}}$---which
changes the sign of $\phi_1$ but does not affect
the lepton doublets $D_\mathrm{L}$---has an important consequence:
the operators $\mathcal{O}_{12}$,
$\mathcal{O}_{21}$,
$\mathcal{O}_{13}$,
and $\mathcal{O}_{31}$ are altogether absent.
The symmetry $\mathbbm{Z}_2^{\mathrm{(tr)}}$,
on the other hand,
changes the sign of $\phi_3$ simultaneously with the interchange
of $D_{\mathrm{L} \mu}$ with $D_{\mathrm{L} \tau}$.
This implies that the coupling matrices $\kk ii\ (i = 1, 2, 3)$
must be of the form
\be
\left( \begin{array}{ccc}
x & y & y \\ y & z & w \\ y & w & z
\end{array} \right),
\label{form1}
\ee
while the matrices $\kk 23$ and $\kk 32 = {\kk 23}^T$ are of the form
\be
\left( \begin{array}{ccc}
0 & p & - p \\ q & s & r \\ - q & - r & - s
\end{array} \right).
\label{form2}
\ee
%


\subsection{The RGE for the $\mathbbm{Z}_2$ and $D_4$ models}

The renormalization group equations for the Yukawa couplings
of the $\mathbbm{Z}_2$ and $D_4$ models are
\ba
16 \pi^2\, \frac{\mathrm{d} y_3}{\mathrm{d} t} &=&
\left( \frac{5}{2} \left| y_3 \right|^2
- \frac{9 g^2 + 15 {g^\prime}^2}{4} \right) y_3,
\\
16 \pi^2\, \frac{\mathrm{d} y_4}{\mathrm{d} t} &=&
\left( \frac{7}{2} \left| y_4 \right|^2
+ \frac{3}{2} \left| y_5 \right|^2
- \frac{9 g^2 + 15 {g^\prime}^2}{4} \right) y_4,
\label{y4rge} \\
16 \pi^2\, \frac{\mathrm{d} y_5}{\mathrm{d} t} &=&
\left( \frac{3}{2} \left| y_4 \right|^2
+ \frac{7}{2} \left| y_5 \right|^2
- \frac{9 g^2 + 15 {g^\prime}^2}{4} \right) y_5.
\label{y5rge}
\ea
The RGE for the scalar-potential couplings are
%
\ba
16 \pi^2\, \frac{\mathrm{d} \lambda_1}{\mathrm{d} t} &=&
24 \lambda_1^2
+ \lambda_4^2 + \left( \lambda_4 + \lambda_7 \right)^2
+ \lambda_5^2 + \left( \lambda_5 + \lambda_8 \right)^2
+ 4 \left| \lambda_{10} \right|^2 + 4 \left| \lambda_{11} \right|^2
\no & &
+ \left( 4 \left| y_3 \right|^2 - C \right) \lambda_1
+ \frac{9 g^4}{8}  + \frac{3 g^2 {g^\prime}^2}{4} + \frac{3 {g^\prime}^4}{8}
- 2 \left| y_3 \right|^4,
\label{1f} \\
16 \pi^2\, \frac{\mathrm{d} \lambda_2}{\mathrm{d} t} &=&
24 \lambda_2^2
+ \lambda_4^2 + \left( \lambda_4 + \lambda_7 \right)^2
+ \lambda_6^2 + \left( \lambda_6 + \lambda_9 \right)^2
+ 4 \left| \lambda_{10} \right|^2 + 4 \left| \lambda_{12} \right|^2
\no & &
+ \left( 8 \left| y_4 \right|^2 - C \right) \lambda_2
+ \frac{9 g^4}{8}  + \frac{3 g^2 {g^\prime}^2}{4} + \frac{3 {g^\prime}^4}{8}
- 4 \left| y_4 \right|^4,
\label{2f} \\
16 \pi^2\, \frac{\mathrm{d} \lambda_3}{\mathrm{d} t} &=&
24 \lambda_3^2
+ \lambda_5^2 + \left( \lambda_5 + \lambda_8 \right)^2
+ \lambda_6^2 + \left( \lambda_6 + \lambda_9 \right)^2
+ 4 \left| \lambda_{11} \right|^2 + 4 \left| \lambda_{12} \right|^2
\no & &
+ \left( 8 \left| y_5 \right|^2 - C \right) \lambda_3
+ \frac{9 g^4}{8}  + \frac{3 g^2 {g^\prime}^2}{4} + \frac{3 {g^\prime}^4}{8}
- 4 \left| y_5 \right|^4,
\label{3f} \\
16 \pi^2\, \frac{\mathrm{d} \lambda_4}{\mathrm{d} t} &=&
\left( \lambda_1 + \lambda_2 \right)
\left( 12 \lambda_4 + 4 \lambda_7 \right)
+ 4 \lambda_4^2 + 2 \lambda_7^2
+ 4 \lambda_5 \lambda_6
+ 2 \left( \lambda_5 \lambda_9 + \lambda_6 \lambda_8 \right)
+ 8 \left| \lambda_{10} \right|^2
\no & &
+ \left( 2 \left| y_3 \right|^2 + 4 \left| y_4 \right|^2 - C
\right) \lambda_4
+ \frac{9 g^4}{4} - \frac{3 g^2 {g^\prime}^2}{2} + \frac{3 {g^\prime}^4}{4},
\label{4f} \\
16 \pi^2\, \frac{\mathrm{d} \lambda_5}{\mathrm{d} t} &=&
\left( \lambda_1 + \lambda_3 \right)
\left( 12 \lambda_5 + 4 \lambda_8 \right)
+ 4 \lambda_5^2 + 2 \lambda_8^2
+ 4 \lambda_4 \lambda_6
+ 2 \left( \lambda_4 \lambda_9 + \lambda_6 \lambda_7 \right)
+ 8 \left| \lambda_{11} \right|^2
\no & &
+ \left( 2 \left| y_3 \right|^2 + 4 \left| y_5 \right|^2 - C
\right) \lambda_5
+ \frac{9 g^4}{4} - \frac{3 g^2 {g^\prime}^2}{2} + \frac{3 {g^\prime}^4}{4},
\label{5f} \\
16 \pi^2\, \frac{\mathrm{d} \lambda_6}{\mathrm{d} t} &=&
\left( \lambda_2 + \lambda_3 \right)
\left( 12 \lambda_6 + 4 \lambda_9 \right)
+ 4 \lambda_6^2 + 2 \lambda_9^2
+ 4 \lambda_4 \lambda_5
+ 2 \left( \lambda_4 \lambda_8 + \lambda_5 \lambda_7 \right)
+ 8 \left| \lambda_{12} \right|^2
\no & &
+ \left( 4 \left| y_4 \right|^2
+ 4 \left| y_5 \right|^2 - C \right) \lambda_6
+ \frac{9 g^4}{4} - \frac{3 g^2 {g^\prime}^2}{2} + \frac{3 {g^\prime}^4}{4}
- 8 \left| y_4 y_5 \right|^2,
\label{6f} \\
16 \pi^2\, \frac{\mathrm{d} \lambda_7}{\mathrm{d} t} &=&
\left( 4 \lambda_1 + 4 \lambda_2 + 8 \lambda_4 + 4 \lambda_7
+ 2 \left| y_3 \right|^2 + 4 \left| y_4 \right|^2 - C \right) \lambda_7
\no & &
+ 2 \lambda_8 \lambda_9 + 32 \left| \lambda_{10} \right|^2
+ 3 g^2 {g^\prime}^2,
\label{7f} \\
16 \pi^2\, \frac{\mathrm{d} \lambda_8}{\mathrm{d} t} &=&
\left( 4 \lambda_1 + 4 \lambda_3 + 8 \lambda_5 + 4 \lambda_8
+ 2 \left| y_3 \right|^2 + 4 \left| y_5 \right|^2 - C \right) \lambda_8
\no & &
+ 2 \lambda_7 \lambda_9 + 32 \left| \lambda_{11} \right|^2
+ 3 g^2 {g^\prime}^2,
\label{8f} \\
16 \pi^2\, \frac{\mathrm{d} \lambda_9}{\mathrm{d} t} &=&
\left( 4 \lambda_2 + 4 \lambda_3 + 8 \lambda_6 + 4 \lambda_9
+ 4 \left| y_4 \right|^2 + 4 \left| y_5 \right|^2 - C
\right) \lambda_9
\no & &
+ 2 \lambda_7 \lambda_8 + 32 \left| \lambda_{12} \right|^2
+ 3 g^2 {g^\prime}^2
- 8 \left| y_4 y_5 \right|^2,
\label{9f} \\
16 \pi^2 \frac{\mathrm{d} \lambda_{10}}{\mathrm{d} t} &=& 
\left( 4 \lambda_1 + 4 \lambda_2 + 8 \lambda_4 + 12 \lambda_7
+ 2 \left| y_3 \right|^2 + 4 \left| y_4 \right|^2
- C \right) \lambda_{10}
\no & &
+ 4 \lambda_{11} \lambda_{12}^\ast,
\label{10f} \\
16 \pi^2 \frac{\mathrm{d} \lambda_{11}}{\mathrm{d} t} &=& 
\left( 4 \lambda_1 + 4 \lambda_3 + 8 \lambda_5
+ 12 \lambda_8 + 2 \left| y_3 \right|^2 + 4 \left| y_5 \right|^2
- C \right) \lambda_{11}
\no & &
+ 4 \lambda_{10} \lambda_{12},
\label{11f} \\
16 \pi^2\, \frac{\mathrm{d} \lambda_{12}}{\mathrm{d} t} &=&
\left( 4 \lambda_2 + 4 \lambda_3 + 8 \lambda_6 + 12 \lambda_9
+ 4 \left| y_4 \right|^2 + 4 \left| y_5 \right|^2
- C \right) \lambda_{12}
\no & &
+ 4 \lambda_{10}^\ast \lambda_{11} - 4 {y_4^\ast}^2 y_5^2,
\label{12f}
\ea
where
\be
C := 9 g^2 + 3 {g^\prime}^2.
\ee
The reason why no fourth-order terms in the Yukawa couplings
and in the gauge couplings appear in the RGE
for $\lambda_{10}$ and $\lambda_{11}$ is that
the condition $\lambda_{10} = \lambda_{11} = 0$
may be enforced through an additional $U(1)$ symmetry:
$\phi_1 \to e^{i\alpha} \phi_1$,
$e_\mathrm{R} \to e^{-i\alpha} e_\mathrm{R}$,
where $\alpha \in \mathbbm{R}$.

We next write down the RGE for the coupling matrices
of the effective neutrino mass operators.
They are
\ba
16 \pi^2\, \frac{\mathrm{d} \kk 11}{\mathrm{d} t} &=&
\left( - 3 g^2 + 4 \lambda_1 + 2 \left| y_3 \right|^2 \right) \kk 11
+ 4 \lambda_{10}^\ast \kk 22 + 4 \lambda_{11}^\ast \kk 33
\no & &
+ \left\{ \kk 11, P - 2 P_1 \right\}, 
\label{beta11} \\
16 \pi^2\, \frac{\mathrm{d} \kk 22}{\mathrm{d} t} &=&
\left( - 3 g^2 + 4 \lambda_2 + 4 \left| y_4 \right|^2 \right) \kk 22
+ 4 \lambda_{10} \kk 11 + 4 \lambda_{12}^\ast \kk 33
\no & &
+ \left\{ \kk 22, P - 2 P_2 \right\}
- 2 \left( \kk 23 P_{23} + P_{23} \kk 32 \right),
\label{beta22} \\
16 \pi^2\, \frac{\mathrm{d} \kk 33}{\mathrm{d} t} &=&
\left( - 3 g^2 + 4 \lambda_3  + 4 \left| y_5 \right|^2 \right) \kk 33
+ 4 \lambda_{11} \kk 11 + 4 \lambda_{12} \kk 22
\no & &
+ \left\{ \kk 33, P - 2 P_3 \right\} 
- 2 \left( \kk 32 P_{32} + P_{32} \kk 23 \right),
\label{beta33} \\
16 \pi^2\, \frac{\mathrm{d} \kk 23}{\mathrm{d} t} &=&
\left( - 3 g^2 + 2 \lambda_6 + 2 \left| y_4 \right|^2
+ 2 \left| y_5 \right|^2 \right) \kk 23
+ 2 \lambda_9 \kk 32 + \left\{ \kk23, P \right\}
\no & &
- 4 \kk 22 P_{32} + 2 P_{32} \kk 22
- 4 P_{23} \kk 33 + 2 \kk 33 P_{23}
\no & &
+ 2 \left[ \kk 23, P_2 - P_3 \right]
- 2 \left( \kk 32 P_3 + P_2 \kk 32 \right),
\label{beta23} \\
16 \pi^2\, \frac{\mathrm{d} \kk 32}{\mathrm{d} t} &=&
\left( - 3 g^2 + 2 \lambda_6 + 2 \left| y_4 \right|^2
+ 2 \left| y_5 \right|^2 \right) \kk 32
+ 2 \lambda_9 \kk 23 + \left\{ \kk32, P \right\}
\no & &
- 4 \kk 33 P_{23} + 2 P_{23} \kk 33
- 4 P_{32} \kk 22 + 2 \kk 22 P_{32}
\no & &
+ 2 \left[ \kk 32, P_3 - P_2 \right]
- 2 \left( \kk 23 P_2 + P_3 \kk 23 \right),
\label{beta32}
\ea
where $\left\{ R, S \right\}$
and $\left[ R, S \right]$ denote the anticommutator and the commutator,
respectively,
of the matrices $R$ and $S$.
Moreover,
we have defined
\ba
P_1 &:=& \mbox{diag} \left( \left| y_3 \right|^2,\
0,\ 0 \right), \\
P_2 &:=& \mbox{diag} \left( 0,\
\left| y_4 \right|^2,\ \left| y_4 \right|^2 \right), \\
P_3 &:=& \mbox{diag} \left( 0,\
\left| y_5 \right|^2,\ \left| y_5 \right|^2 \right), \\
P_{23} &:=& \mbox{diag} \left( 0,\ y_4 y_5^\ast,\ - y_4 y_5^\ast \right),
\label{P23} \\
P_{32} &:=& \mbox{diag} \left( 0,\ y_4^\ast y_5,\ - y_4^\ast y_5 \right).
\label{P32}
\ea
Notice that the matrix $P$ in~(\ref{Pdefinition})
is equal to $\left( P_1 + P_2 + P_3 \right) / 2$.


\section{Predictions of the $\mathbbm{Z}_2$ and $D_4$ models}
\label{sectionpredictions}

The Lagrangian of neutrino Majorana masses is
\be
\mathcal{L}_\mathrm{Majorana} = \frac{1}{2}
\sum_{\alpha, \beta = e, \mu, \tau}
\nu_{\mathrm{L} \alpha}^T C^{-1}
\left( \mathcal{M}_\nu \right) _{\alpha \beta} \nu_{\mathrm{L} \beta}
+ \mathrm{H.c.},
\ee
where $\mathcal{M}_\nu = \mathcal{M}_\nu^T$.
Taking $b = d = 2$ in~(\ref{Op}),
it is clear that
\be
\mathcal{O}_{ij} = \sum_{\alpha, \beta = e, \mu, \tau}
\kk ij_{\alpha \beta} \phi_i^0 \phi_j^0
\nu_{\mathrm{L} \alpha}^T C^{-1} \nu_{\mathrm{L} \beta} + \cdots.
\ee
Therefore,
if we denote the VEV of $\phi_i^0$ by $v_i$,
then the neutrino Majorana mass matrix $\mathcal{M}_\nu$ is given by
\be
\frac{1}{2}\, \mathcal{M}_\nu
= \sum_{i=1}^3 v_i^2 \kk ii + v_2 v_3 \left[ \kk 23 + \kk 32 \right],
\label{mass}
\ee
since $\kk 12 = \kk 21 = \kk 13 = \kk 31 = 0$
in the $\mathbbm{Z}_2$ and $D_4$ models.
This is valid at all scales $t$.

In general one may write~\cite{GJKLST}
\be
\mathcal{M}_\nu = \left( \matrix{
X & A \left( 1 + \e \right) & A \left( 1 - \e \right) \cr
A \left( 1 + \e \right) & B \left( 1 + \e^\prime \right) & C \cr
A \left( 1 - \e \right) & C & B \left( 1 - \e^\prime \right) \cr}
\right).
\label{mnuform}
\ee
We already know that
the form of the flavour coupling matrices $\kk ii$
is described by~(\ref{form1}),
while $\kk 23 = {\kk 32}^T$ is decribed by~(\ref{form2}).
Therefore,
\be
\sum_{i=1}^3 2 v_i^2 \kk ii = \left( \matrix{
X & A & A \cr A & B & C \cr A & C & B \cr}
\right),
\label{sum}
\ee
while
\be
\kappa^{(c)} := \kk 23 + \kk 32 = \left( \begin{array}{ccc}
0 & c_1 & - c_1 \\ c_1 & c_2 & 0 \\ - c_1 & 0 & - c_2
\end{array} \right),
\label{kappac}
\ee
with
\ba
\e A &=& 2 v_2 v_3 c_1,
\label{ea}
\\
\e^\prime B &=& 2 v_2 v_3 c_2.
\ea
Once again,
all this is valid at any scale $t$.

In both the $\mathbbm{Z}_2$ and $D_4$ models,
the symmetry $\mathbbm{Z}_2^{\mathrm{(aux)}}$
inverts the signs of the right-handed-neutrino fields
which are present above the seesaw scale.
Hence,
only the doublet $\phi_1$ has Yukawa couplings
to those fields,
above the high scale.
This implies that $\mathcal{M}_\nu \left( t_0 \right)$,
where $t_0 := \ln{m_R}$,
originates solely from the VEV of $\phi_1^0$~\cite{GLZ2,GLD4}.
Therefore,
at the seesaw scale~\cite{softD4}
\be
\begin{array}{rcl}
\kk 11 \left( t_0 \right) &=&
\mathcal{M}_\nu \left( t_0 \right) / \left( 2 v_1^2 \right), \\*[1mm]
\kk ij \left( t_0 \right) &=& 0
\ \mbox{for all other}\ (ij).
\end{array}
\label{initial}
\ee
We conclude that $\mathcal{M}_\nu \left( t_0 \right)$
has the same form as $\kk 11$,
i.e.\ $\mathcal{M}_\nu \left( t_0 \right)$ is of the form~(\ref{form1}).
Clearly then,
$\left( 0, 1, -1 \right)^T$
is an eigenvector of $\mathcal{M}_\nu \left( t_0 \right)$ and therefore,
at the seesaw scale,
the predictions~(\ref{angles}) hold.

At any other scale,
though,
the matrix $\kappa^{(c)}$ in~(\ref{kappac}) is not zero.
Thus,
for any $t < t_0$,
$\mathcal{M}_\nu \left( t \right)$ is not
$\mu \leftrightarrow \tau$-symmetric.
This fact renders the predictions~(\ref{angles})
\emph{inexact for any scale other than the seesaw scale}.
In~\cite{GJKLST} it has been shown that,
if one assumes the parameters $\e$ and $\e^\prime$
in~(\ref{mnuform}) to be small,
then,
to first order in those parameters,
one has,
instead of~(\ref{angles}),
\ba
\label{spue3}
U_{e3} &=&
\frac{s_{12} c_{12}}{m_3^2 - m_2^2} \left(
\bar \e s_{12}^2 \hat m_2^\ast
+ \bar \e^\ast s_{12}^2 m_3
- \bar \e^\prime \hat m_2^\ast
- \bar \e^{\prime \ast} m_3
\right)
\nonumber \\ & &
+ \frac{s_{12} c_{12}}{m_3^2 - m_1^2} \left(
\bar \e c_{12}^2 \hat m_1^\ast + \bar \e^\ast c_{12}^2 m_3
+ \bar \e^\prime \hat m_1^\ast
+ \bar \e^{\prime \ast} m_3
\right),
\\
\label{spcos}
\cos{2 \theta_{23}} &=&
{\rm Re} \left[
\frac{2 c_{12}^2}{m_3^2-m_2^2}
\left( \bar \e s_{12}^2 - \bar \e^\prime \right)
\left( \hat m_2 + m_3 \right)^\ast
\right. \no & & \left.
- \frac{2 s_{12}^2}{m_3^2 - m_1^2}
\left( \bar \e c_{12}^2 + \bar \e^\prime \right)
\left( \hat m_1 + m_3 \right)^\ast
\right],
\ea
where
\ba
\bar \e &:=& \left( \hat m_1 - \hat m_2 \right) \e,
\\
\bar \e^\prime &:=& \frac
{\hat m_1 s_{12}^2 + \hat m_2 c_{12}^2 + m_3}{2}\, \e^\prime,
\label{e'}
\ea
and
\ba
\hat m_1 &=& m_1 e^{- 2 i \rho},
\\
\hat m_2 &=& m_2 e^{- 2 i \sigma}.
\ea
Here,
$m_1$,
$m_2$,
and $m_3$ are the (real, non-negative) neutrino masses,
while $\rho$ and $\sigma$ are Majorana phases.
In~(\ref{spue3})--(\ref{e'}),
$s_{12}$ and $c_{12}$ are the sine and cosine,
respectively,
of the solar mixing angle.
We are using the standard parametrization~\cite{maltoni,altarelli}
for the lepton mixing matrix $U$;
that parametrization fixes the significance of the phase of~(\ref{spue3}).

We see in~(\ref{spue3}) and~(\ref{spcos})
that the deviation from the predictions~(\ref{angles})
is numerically of the same order
as the (small) parameters $\e$ and $\e^\prime$,
except in the case of quasi-degenerate neutrinos
with common mass $m_0$;
in that case,
an enhancement factor 
$m_0^2 / \Delta m^2_\mathrm{atm}$ appears,
where $\Delta m^2_\mathrm{atm} = \left| m_3^2 - m_2^2 \right|
\approx \left| m_3^2 - m_1^2 \right|$
is the atmospheric mass-squared difference.


\section{An approximate solution of the RGE}
\label{sectionupper}

In this section we assume the Yukawa couplings $y_3$,
$y_4$,
and $y_5$ to be very small.
This assumption is reasonable when one considers
the $\mathbbm{Z}_2$ or $D_4$ models with their minimal content
of three Higgs doublets,
since in that case $\sum_{i=1}^3 \left| v_i \right|^2$
must be equal to $\left( 174\, \mathrm{GeV} \right)^2$,
and therefore all the $\left| v_i \right|$ are in principle
much larger than the charged-lepton masses.
We estimate,
to first order in the Yukawa couplings squared,
the deviation of $\mathcal{M}_\nu \left( t_1 \right)$,
where $t_1 := \ln{m_Z}$,
from the $\mu \leftrightarrow \tau$-symmetric form.
We thus find out the likely magnitudes
of $U_{e3}$ and $\cos{2 \theta_{23}}$ at the electroweak scale.

The starting point is~(\ref{beta11})--(\ref{beta32}).
We drop all the Yukawa couplings from~(\ref{beta11})--(\ref{beta33});
in~(\ref{beta23}) and~(\ref{beta32}),
on the other hand,
we keep the terms with $P_{23}$ and $P_{32}$,
since it is those terms which induce corrections
to the $\mu \leftrightarrow \tau$-symmetric 
form of $\mathcal{M}_\nu \left( t_0 \right)$.
We thus obtain
\ba
16 \pi^2\, \frac{\mathrm{d} \kk 11}{\mathrm{d} t} &\approx&
\left( - 3 g^2 + 4 \lambda_1 \right) \kk 11
+ 4 \lambda_{10}^\ast \kk 22 + 4 \lambda_{11}^\ast \kk 33,
\label{k1} \\
16 \pi^2\, \frac{\mathrm{d} \kk 22}{\mathrm{d} t} &\approx&
\left( - 3 g^2 + 4 \lambda_2 \right) \kk 22
+ 4 \lambda_{10} \kk 11 + 4 \lambda_{12}^\ast \kk 33,
\label{k2} \\
16 \pi^2\, \frac{\mathrm{d} \kk 33}{\mathrm{d} t} &\approx&
\left( - 3 g^2 + 4 \lambda_3 \right) \kk 33
+ 4 \lambda_{11} \kk 11 + 4 \lambda_{12} \kk 22,
\label{k3}
\ea
and
\be
16 \pi^2\, \frac{\mathrm{d} \kappa^{(c)}}{\mathrm{d} t} = 
\left( -3 g^2 + 2 \lambda_6 + 2 \lambda_9 \right) \kappa^{(c)}
- 2 \left\{ P_{32}, \kappa^{(22)} \right\}
- 2 \left\{ P_{23}, \kappa^{(33)} \right\},
\label{2332}
\ee
where $\kappa^{(c)} = \kk 23 + \kk 32$ as in~(\ref{kappac}).
Using~(\ref{P23}) and~(\ref{P32}),
we obtain from~(\ref{2332}) that
\ba
16 \pi^2\, \frac{\mathrm{d} c_1}{\mathrm{d} t} &=&
\left( - 3 g^2 + 2 \lambda_6 + 2 \lambda_9 \right) c_1 -
2 \left[ y_4^\ast y_5 \kk 22_{12} + y_4 y_5^\ast \kk 33_{12} \right],
\label{c1eq} \\
16 \pi^2\, \frac{\mathrm{d} c_2}{\mathrm{d} t} &=&
\left( - 3 g^2 + 2 \lambda_6 + 2 \lambda_9 \right) c_2 -
4 \left[ y_4^\ast y_5 \kk 22_{22} + y_4 y_5^\ast \kk 33_{22} \right].
\label{c2eq}
\ea

Formally we can write the solution
of the coupled differential equations~(\ref{k1})--(\ref{k3}) as
\be
\kk ii \left( t \right) = \sum_{j=1}^3
T \left( t, t^\prime \right)_{ij} \kk jj \left( t^\prime \right).
\ee
\emph{All} the matrix elements of $\kk ii$
evolve according to the same operator $T \left( t, t^\prime \right)$.
Since at the scale $t_0$ only $\kk 11$ is non-vanishing,
\be
\kk ii \left( t \right) =
T \left( t, t_0 \right)_{i1} \kk 11 \left( t_0 \right).
\label{ttt}
\ee

Formally,
the solutions to~(\ref{c1eq}) and~(\ref{c2eq}) are easily written down.
Defining
\be
S_c \left( t \right) :=
\exp \left[ \frac{1}{16 \pi^2} \int_{t_0}^t \mathrm{d} t^\prime 
\left( -3 g^2 + 2 \lambda_6 + 2 \lambda_9 \right)
\left( t^\prime \right) \right]
\label{sc}
\ee
and taking into account that 
$c_1 \left( t_0 \right) = c_2 \left( t_0 \right) = 0$,
one has
\ba
c_1 \left( t \right) &=&
- \frac{1}{8 \pi^2}\, S_c \left( t \right) 
\int_{t_0}^t \mathrm{d} t^\prime\,
S_c^{-1} \left( t^\prime \right)
\left[ y_4^\ast y_5 \kk 22_{12}
+ y_4 y_5^\ast \kk 33_{12} \right] \left( t^\prime \right)
\no &=&
- \frac{\kk 11_{12} \left( t_0 \right)}{8 \pi^2}\, S_c \left( t \right) 
\int_{t_0}^t \mathrm{d} t^\prime\,
S_c^{-1} \left( t^\prime \right)
\left[ \nu \left( t^\prime \right) T \left( t^\prime, t_0 \right)_{21}
+ \nu^\ast \left( t^\prime \right) T \left( t^\prime, t_0 \right)_{31}
\right],
\label{c1} \\
c_2 \left( t \right) &=&
- \frac{1}{4 \pi^2}\, S_c \left( t \right) 
\int_{t_0}^t \mathrm{d} t^\prime\,
S_c^{-1} \left( t^\prime \right)
\left[ y_4^\ast y_5 \kk 22_{22}
+ y_4 y_5^\ast \kk 33_{22} \right] \left( t^\prime \right)
\no &=&
- \frac{\kk 11_{22} \left( t_0 \right)}{4 \pi^2}\, S_c \left( t \right) 
\int_{t_0}^t \mathrm{d} t^\prime\,
S_c^{-1} \left( t^\prime \right)
\left[ \nu \left( t^\prime \right) T \left( t^\prime, t_0 \right)_{21}
+ \nu^\ast \left( t^\prime \right) T \left( t^\prime, t_0 \right)_{31}
\right],
\label{c2}
\ea
where we have used~(\ref{ttt}) and defined $\nu := y_4^\ast y_5$.

We now make use of~(\ref{sum}) and~(\ref{ea}) to write
\ba
\e \left( t_1 \right) &=&
\frac{v_2 v_3 c_1 \left( t_1 \right)}
{\sum_{i=1}^3 v_i^2 \kk ii_{12} \left( t_1 \right)}
\no &=&
\frac{v_2 v_3 c_1 \left( t_1 \right)}
{\sum_{i=1}^3 v_i^2 T \left( t_1, t_0 \right)_{i1}}\,
\frac{1}{\kk 11_{12} \left( t_0 \right)}.
\label{epsilon}
\ea
Similarly,
\be
\e^\prime \left( t_1 \right) =
\frac{v_2 v_3 c_2 \left( t_1 \right)}
{\sum_{i=1}^3 v_i^2 T \left( t_1, t_0 \right)_{i1}}\,
\frac{1}{\kk 11_{22} \left( t_0 \right)}.
\label{epsilonprime}
\ee
Putting~(\ref{c1})--(\ref{epsilonprime}) together,
we conclude that
\be
\e \left( t_1 \right) =
- \frac{v_2 v_3}
{8 \pi^2 \sum_{i=1}^3 v_i^2 T \left( t_1, t_0 \right)_{i1}}\,
S_c \left( t_1 \right)
\int_{t_0}^{t_1} \mathrm{d} t^\prime\,
S_c^{-1} \left( t^\prime \right)
\left[ \nu \left( t^\prime \right) T \left( t^\prime, t_0 \right)_{21}
+ \nu^\ast \left( t^\prime \right) T \left( t^\prime, t_0 \right)_{31}
\right]
\label{finale}
\ee
while
$\e^\prime \left( t_1 \right) = 2 \e \left( t_1 \right)$.
From~(\ref{spue3}) and~(\ref{spcos}) one can derive that 
\be
\label{cosU}
\e^\prime = 2 \e \ \Rightarrow \
\left\{ \begin{array}{rcl}
U_{e3} &=& 2 m_3 c_{12} s_{12} \left(
{\displaystyle \frac{\hat m_1^\ast + m_3}{m_3^2 - m_1^2}} + 
{\displaystyle \frac{\hat m_2^\ast + m_3}{m_2^2 - m_3^2}}
\right) \mbox{Re}\, \epsilon,
\\[4mm]
\cos{2 \theta_{23}} &=& 
2 \left(
s_{12}^2 {\displaystyle \frac{\left| \hat m_1 + m_3 \right|^2}
{m_1^2 - m_3^2}} + 
c_{12}^2 {\displaystyle \frac{\left| \hat m_2 + m_3 \right|^2}
{m_2^2 - m_3^2}} \right)
\mbox{Re}\, \epsilon.
\end{array}
\right.
\ee

Now we want to estimate the maximum possible
order of magnitude of $\e \left( t_1 \right)$
by using~(\ref{finale}). 
The length of the integration interval of $t^\prime$
is $t_0 - t_1 = \ln{\left( m_R/m_Z \right)} \sim 10$.
The functions $S_c \left( t \right)$
and $S_c^{-1} \left( t^\prime \right)$ are of order 1 since,
in~(\ref{sc}),
$\lambda_6 / \left( 16 \pi^2 \right)$
and $\lambda_9 / \left( 16 \pi^2 \right)$ are necessarily small.
The functions $T \left( t^\prime, t_0 \right)_{21}$
and $T \left( t^\prime, t_0 \right)_{31}$
are governed by $\lambda_{10}$ and $\lambda_{11}$;
in any case,
$T \left( t^\prime, t_0 \right)_{21} / T \left( t_1, t_0 \right)_{i1}$
and
$T \left( t^\prime, t_0 \right)_{31} / T \left( t_1, t_0 \right)_{i1}$
should be $\lesssim 1$.
Similarly,
$v_2 v_3 / v_i^2$ should not be larger than 1.
We conclude that
\be
\left| \e \left( t_1 \right) \right| 
\sim \frac{10 | \nu |}{8 \pi^2} 
\sim \frac{\left| y_4 y_5 \right|}{10}.
\ee
Even if we allow for rather small VEVs,
$\left| y_{4,5} \right|$ cannot be larger than $0.1$.
We thus have the generous upper bound
$\left| \e \left( t_1 \right) \right| \lesssim 10^{-3}$.

Equation~(\ref{cosU}) tells us that the only chance
to have a non-negligible $U_{e3}$
is in the case of a degenerate neutrino spectrum.
Let us consider the extreme case
of a common mass
$m_0 = 0.3\, \mathrm{eV}$~\cite{barger}.
Since $\Delta m^2_{\mathrm{atm}} \simeq 2 \times 10^{-3}\, \mathrm{eV}^2$,
in that case we have
$| U_{e3} | \simeq 100 \left | \mathrm{Re}\,\epsilon \right|
\lesssim 0.1$.
Here we have used Majorana phases
$\rho = \sigma = 0$ for simplicity.
That choice of $m_0$ is indeed extreme;
if take $m_0 = 0.1\, \mathrm{eV}$ instead,
then the upper bound becomes one order of magnitude smaller,
due to the quadratic dependence of $\left| U_{e3} \right|$
on $m_0$ in the case of a degenerate neutrino spectrum.
In any case,
we expect $\left| U_{e3} \right|$
and $\left| \cos{2 \theta_{23}} \right|$
to be no larger than $0.1$ in our model,
but most likely they are two or more orders of magnitude smaller.
 

\section{Summary} \label{summary}

In this paper we have computed the RGE
for the dimension-five neutrino mass operators
in the multi-Higgs-doublet SM.
Thus, the main result of this paper is~(\ref{RGEkappa}),
which describes the evolution of the coupling matrices
of the mass operators
in the SM with an arbitrary number of Higgs doublets. 
We have argued in favour of the usefulness of~(\ref{RGEkappa})
by citing models
for lepton mixing
which have been constructed in the framework of the multi-Higgs-doublet SM. 

As an application of our RGE
we have considered the $\mathbbm{Z}_2$ model of~\cite{GLZ2}
and the $D_4$ model of~\cite{GLD4},
which---from the field-theoretical point of view---are identical
below the seesaw scale.
The predictions~(\ref{angles}) of those models
hold at the seesaw scale and we have used the RGE
to estimate the corrections to those predictions
which appear due to the evolution of the coupling matrices
of the dimension-five neutrino mass operators
down to the electroweak scale.
We have found that those corrections are in general negligible,
with the possible exception of a degenerate neutrino mass spectrum
with a rather large common mass $m_0 \gtrsim 0.2\, \mathrm{eV}$.
In that case,
$s^2_{13}$ could be as large as 0.01
and be within the sensitivity
of the planned long-baseline neutrino experiments~\cite{maltoni}.
On the other hand,
even in the degenerate neutrino case
the deviation of $\theta_{23}$ from
$\pi / 4$ will be hard to uncover
in those experiments~\cite{antusch},
since they will be sensitive to the parameter $\sin^2{2 \theta_{23}}
= 1 - \cos^2{2 \theta_{23}}$ and we have estimated  
$\cos^2{2 \theta_{23}}\, \lesssim \, 0.01$ in our models.
Thus,
with respect to the experiments presently envisaged
the models discussed here have the following properties:
should a non-zero $s^2_{13}$ be discovered,
then the neutrino mass spectrum must be degenerate;
while deviations from $\sin^2{2 \theta_{23}} = 1$ should be invisible.


\appendix
\setcounter{equation}{0}
\renewcommand{\theequation}{A\arabic{equation}}

\section{Vertex corrections to the neutrino mass operator}

The last two lines of~(\ref{RGEkappa})
originate in vertex corrections of the type displayed
in figure~\ref{fig}.
In this appendix we show how we have arrived at those two lines.
Perturbation theory yields the expression 
\ba
& &
\frac{(-i)^2}{2!}\, \mathbf{T} \left[
\sum_{m,n = 1}^{n_H}
\sum_{a, b, c, d = 1}^2
\left( D_{\mathrm{L}a}^T \kk mn C^{-1} D_{\mathrm{L}c}\,
\varepsilon^{ab} \phi_{mb}\,
\varepsilon^{cd} \phi_{nd}
\right)_x
\right. \no & & \left.
\times\, 2
\int \! \mathrm{d}^d x_1
\sum_{k=1}^{n_H} \sum_{e=1}^2
\left( \bar D_{\mathrm{L}e} Y_k^\dagger \phi_{ke}
\ell_\mathrm{R} \right)_{x_1} 
\int \! \mathrm{d}^d x_2 
\sum_{l=1}^{n_H} \sum_{f=1}^2
\left( \bar \ell_\mathrm{R} \phi_{lf}^\dagger Y_l
D_{\mathrm{L}f} \right)_{x_2}
\right],
\label{second}
\ea
where $d$ is the dimension of space--time and $x$,
$x_1$,
$x_2$ are space--time points.
We have left out the flavour indices in~(\ref{second}).
The gauge-$SU(2)$ indices are $a, b, \ldots, f$.
The symbol $\mathbf{T}$ denotes time ordering.
When computing~(\ref{second}),
the field $\ell_\mathrm{R}$ must be contracted
with $\bar \ell_\mathrm{R}$.
As for $\bar D_{\mathrm{L}e}$,
it may be contracted either with $D_{\mathrm{L}a}$
or with $D_{\mathrm{L}c}$;
it is easy to see that both possibilities
yield the same contribution to the RGE of $\kk ij$---this fact explains
the factor 2 in the second line of~(\ref{RGEkappa}).
In the following we compute explicitly the case where
$\bar D_{\mathrm{L}e}$ is contracted with $D_{\mathrm{L}c}$.

For the contraction of $\phi_{lf}^\dagger$
there are also two possibilities:
one may contract it either with $\phi_{mb}$ or with $\phi_{nd}$.
We consider the second possibility first.
We use dimensional regularization with minimal subtraction.
In the evaluation of~(\ref{second})
we only need the pole terms in $\epsilon = 4 - d$.
The computation is straightforward and we arrive at 
\be
\frac{1}{16 \pi^2 \epsilon}\,
\sum_{k,m,n = 1}^{n_H}
\sum_{a,b,c,d = 1}^2
D_{\mathrm{L}a}^T C^{-1}
\left[ \kk mn Y_k^\dagger Y_n \right] D_{\mathrm{L}d}\,
\varepsilon^{ab} \phi_{mb}\,
\varepsilon^{dc} \phi_{kc},
\label{1}
\ee
where all the fields are now meant to be
at the same space--time point.
The minus sign from the factor $(-i)^2$
in~(\ref{second}) has been removed
through the interchange of the indices $c$ and $d$
in $\varepsilon^{cd}$.
If,
instead,
we contract $\phi_{lf}^\dagger$ with $\phi_{mb}$,
then we get
\be
- \frac{1}{16 \pi^2 \epsilon}\,
\sum_{k,m,n = 1}^{n_H}
\sum_{a,b,c,d = 1}^2
D_{\mathrm{L}a}^T C^{-1} \left[ \kk mn Y_k^\dagger Y_m \right]
D_{\mathrm{L}b}\,
\varepsilon^{ab} \varepsilon^{cd} \phi_{kc} \phi_{nd}.
\label{2}
\ee
The $SU(2)$ structure of~(\ref{2})
is different from the one in~(\ref{Op}).
Therefore,
we need to apply the identity
\be
\varepsilon^{ab} \varepsilon^{cd}
+ \varepsilon^{ac} \varepsilon^{db}
+ \varepsilon^{ad} \varepsilon^{bc} = 0
\label{epsrel}
\ee
to~(\ref{2}),
obtaining
\be
\frac{1}{16 \pi^2 \epsilon}\,
\sum_{k,m,n = 1}^{n_H}
\sum_{a,b,c,d = 1}^2
D_{\mathrm{L}a}^T C^{-1} \left[ \kk mn Y_k^\dagger Y_m \right]
D_{\mathrm{L}b}
\left( - \varepsilon^{ac} \phi_{kc}\, \varepsilon^{bd} \phi_{nd}
+ \varepsilon^{ad} \phi_{nd}\, \varepsilon^{bc} \phi_{kc} \right).
\label{2'}
\ee
The terms in the last two lines of~(\ref{RGEkappa})
are obtained from~(\ref{1}) and~(\ref{2'}) as follows.
Firstly,
one substitutes the factor $1 / \epsilon$
by $-1$~\cite{drees2}.
Secondly,
the indices $k$ and $m$---in~(\ref{1})---or $k$
and $n$---in~(\ref{2'})---of the scalar doublets
must be replaced by $i$ and $j$;
the contribution to the beta function of $\kk ij$
is then given by the flavour matrix in between the lepton doublets.
If,
in each expression,
the first Higgs doublet is labeled $i$ and the second one is labeled $j$,
then one obtains the terms in the second line of~(\ref{RGEkappa}):
$- \sum_ n \kk in Y_j^\dagger Y_n$ from~(\ref{1}),
$\sum_m \left( \kk mj Y_i^\dagger Y_m
- \sum_m \kk mi Y_j^\dagger Y_m \right)$ from~(\ref{2'}).
Reversing the role of $i$ and $j$ 
leads to the terms in the last line of~(\ref{RGEkappa}).

\vskip 1 cm
\noindent \textbf{Acknowledgements}:
The work of L.L.\ has been supported
by the Portuguese \textit{Funda\c c\~ao para a Ci\^encia e a Tecnologia}
under the project U777--Plurianual.

\end{document}